\documentclass[12pt]{iopart}

\usepackage{iopams}
\usepackage{graphicx}

\begin{document}

\title[Yield statistics of interpolated superoscillations]{Yield statistics of interpolated superoscillations}

\author{Eytan Katzav$^1$, Ehud Perlsman$^2$ and Moshe Schwartz$^{2,3}$}

\address{$^1$Racah Institute of Physics, The Hebrew University, Jerusalem, 91904, Israel.}
\address{$^2$Department of Physics, Raymond and Beverly Sackler Faculty  Exact Sciences, Tel Aviv University, Tel Aviv 69978, Israel.}
\address{$^3$Faculty of Engineering, H.I.T. - Holon Institute of Technology, Holon 5810201, Israel.}
\eads{\mailto{eytan.katzav@mail.huji.ac.il}, \mailto{udikhamudi@gmail.com}, \mailto{bricki@netvision.net.il}}

\begin{abstract}

Yield Optimized Interpolated Superoscillations (YOIS) have been recently introduced as a means for possibly making the use of the phenomenon of superoscillation practical. In this paper we study how good is a superoscillation that is not optimal. Namely, by how much is the yield decreased when the signal departs from the optimal one. We consider two situations. One is the case where the signal strictly obeys the interpolation requirement and the other is when that requirement is relaxed. In the latter case the yield can be increased at the expense of deterioration of signal quality. An important conclusion is that optimizing superoscillations may be challenging in terms of the precision needed, however, storing and using them is not at all that sensitive. This is of great importance in any physical system where noise and error are inevitable.

\end{abstract}

\vspace{2pc}

\submitto{\JPA (\today)}
\date
\maketitle

\section{Introduction}

Superoscillatory signals are band limited functions that oscillate over some regions with a frequency larger than that of its maximal Fourier component. A number of examples have been given in the past for such functions with very interesting applications in quantum mechanics \cite{Aharonov88,Aharonov90,Berry94a,Berry94b,Kempf2000,Berry06,Aharonov11,Aharonov15}, signal processing \cite{Slepian61,Levi65,Slepian78,Kempf02,Kempf06,Lee14a,Lee14b} and in optics, where superoscillations are intimately related to super-resolution \cite{Zheludov08,Zheludov09,Zalevsky11} with some recent exciting experimental achievements \cite{Yuan14,Rogers12}. For a recent review see \cite{Lindberg12}.

A key issue in the field is quantifying the superoscillations using some local frequency, for example by taking the local rate of change of the phase of the signal \cite{Berry14} among other options. Another question of much interest is how superoscillations interact with physical matter during propagation \cite{Baranov14,Eliezer14}, with some indication (both theoretically \cite{Baranov15} and experimentally \cite{Remez15}) that nonlinearities may actually play an important role in utilizing them. Furthermore, ref. \cite{Barnett13} provides a prediction of a quantum effect directly related to superoscillations, causing momentum 'superkicks' imparted to a small particle located near optical vortices.

However, although the phenomenon of superoscillation suggests many practical applications, its uses are presently rather limited, because the superoscillatory signal exist in limited intervals and that the amplitude of the superoscillations in those regions is extremely small compared to typical values of the amplitude in non-superoscillating regions \cite{Aharonov90,Kempf02,Kempf06}. This often renders algorithms for generating them numerically unstable \cite{Lee14c}. One of the methods for obtaining superoscillations that lends itself easily to improving the situation described above, is that of forcing an interpolation within the prescribed interval that will necessarily lead to the required superoscillation \cite{Kempf06,Katzav13}. Actually, fitting functions to prescribed forms inducing superoscillations by interpolation has been applied to optical superresolution in \cite{Makris2011,Greenfield2013}, and the resulting matrix inversion problems were explored in \cite{Berry2013b}.

We start our discussion by explaining the concept of interpolated superoscillation \cite{Kempf06,Katzav13}. Consider the signal 
\begin{equation}
f(t) = \frac{A_0}{\sqrt{2\pi}} + \sum\limits_{m = 1}^N {\frac{A_m}{\sqrt \pi}} \cos (mt) ,
\label{eq:1}
\end{equation}
which is periodic with a period of $2 \pi$. Choose a sub-interval $\left[0,a\right]$ with $a < \pi $ and impose on the function $f(t)$ $M$ constraints in the interval, such that $f(t_j) = \mu_j$ for $0 \le t_j \le a$ and $j = 0, \ldots ,M - 1$ The constraints result in a set of $M$ linear equations in $N + 1$ unknowns of the form,
\begin{equation}
\sum\limits_{m = 0}^N C_{jm} A_m \equiv {\bf{C}}_j \cdot {\bf A} = \mu_j \, ,
\label{eq:2}
\end{equation}
where 
\begin{equation}
C_{jm} = \frac{1}{\sqrt \pi}
\left\{ \begin{array}{l}
\cos (m{t_j})      \hspace{10mm} m \ne 0\\
\frac{1}{\sqrt 2}  \hspace{21mm} m = 0
\end{array} \right. .
\end{equation}

\noindent
Generically this set of equations: has no solution for $M > N + 1$, has one solution for $M = N + 1$ and a whole space of solutions for $M < N + 1$. A particular choice of great interest \cite{Kempf06,Katzav13} is
\begin{equation}
t_j=\frac{a}{M-1}j \qquad \rm{and} \qquad \mu_j=(-1)^j \, .
\label{eq:3}
\end{equation}
Provided $M \le N + 1$, this choice constrains the function to oscillate within the interval $[ - a,a]$ between the values $ \pm 1$ with a frequency 
\begin{equation}
\omega  = \frac{\pi (M-1)}{a} \, .
\label{eq:4}
\end{equation}
It is thus clear that the frequency of oscillation within the interval $[-a,a]$ can be increased indefinitely regardless of the largest frequency $\cos (Nt)$ included in the signal (\ref{eq:1}). As is well known, this comes, at a cost. The energy in the superoscillatory region is extremely small relative to the total energy. Therefore optimization is required if the use of the phenomenon of superoscillation is to become practical.

A natural step is to optimize the superoscillating function for fixed $a$ and $M < N + 1$. However, one has to decide first in what sense to optimize it. One approach \cite{Kempf06,Katzav13} is to consider the energy of the signal, $E = \int\limits_{-\infty}^\infty  f^2(t)dt$, use the fact that $f$ is band limited and minimize the energy under the interpolation constraints. In the periodic function case, the equivalent would be to minimize, $E = \int\limits_{-\pi}^\pi  f^2(t)dt$  under the choice of the Fourier coefficients given by equation (\ref{eq:1}) and the constraints (\ref{eq:3}) . A step forward would be to maximize the superoscillation yield, defined as
\begin{equation}
Y(M,a) = \frac{\int\limits_{-a}^a {f^2(t)dt}}{\int\limits_{-\infty }^\infty {f^2(t)dt}} \, ,
\label{eq:5}
\end{equation}
rather than the total energy. As will become evident in the following the two methods of optimization are very close to each other, provided the optimal yield is indeed very small, which in many situations is the case. 
It is in place to say that defining the yield in terms of the energy of the signal is one option, which relies on the function being square integrable, foe example. However, square integrability is not always relevant because it can be related to features of the signal far away from the superoscillatory region, such as a slow decay of the tail, see \cite{Berry2013a} for such an example. Therefore, more generally, one can devise other definitions of the superoscillatory yield which are more localized around the superoscillatory region.

Plugging a signal described by its Fourier components, as in Eq. (\ref{eq:1}), into the definition of the yield, given by Eq. (\ref{eq:5}), we can obtain the following expression for the yield

\begin{equation}
Y(N,M,a) = \frac{\sum\limits_{m,n = 0}^N {\Delta_{mn} A_m A_n}}{\sum\limits_{m = 0}^N A_m^2} \equiv \frac{I}{D} \, ,
\label{eq:6}
\end{equation}

\noindent
where the entries of the matrix $\Delta $ are given by 
\begin{equation}
\Delta_{mn} \equiv \left\{ \begin{array}{l}
\frac{2}{\pi }\frac{m\cos (na)\sin (ma) - n\cos (ma)\sin (na)}{m^2-n^2} \qquad m \ne n \ne 0\\
\frac{1}{\pi }\left( a + \frac{\sin (2na)}{2n} \right) \hspace{33mm} m=n \ne 0\\
\frac{\sqrt{2}}{\pi n}\sin (n a) \hspace{42.5mm} m=0, n \ne 0\\
\frac{\sqrt{2}}{\pi m}\sin (m a) \hspace{41mm} n=0, m \ne 0\\
\frac{a}{\pi}                    \hspace{58mm} m=n=0
\end{array} \right. .
\label{eq:7}
\end{equation}

\noindent
To implement the optimization of the yield \cite{Katzav13}, the original degrees of freedom, are rotated, going from the original set of the $A$'s to a new set of degrees of freedom $B$, as to break the linear $N + 1$ dimensional space, into a direct sum of two subspaces. The first $M$ dimensional subspace is the space, $S_M$, where the $M$ coefficients $\{B_n\}$ are independently constrained and attain prescribed values determined by the constraints. Thus, from all that subspace, we choose a specific vector ${\bf{B}}_M^C$. The remaining $(N + 1 - M)$-dimensional subspace, ${S_{N + 1 - M}}$, is totally unconstrained. 

\noindent
The set of interpolating signals is defined thus by
\begin{equation}
{\bf{B}} = {{\bf{B}}_{N + 1 - M}} \oplus {\bf{B}}_M^C \, ,
\label{eq:8}
\end{equation}
where ${\bf{B}_{N + 1 - M}} \in {S_{N + 1 - M}}$.

In Ref. \cite{Katzav13} we have optimized the yield under the interpolation constraints. In this paper we explore the effect of noise on the superoscillatory signal. Such noise is abundant in many physical systems and must be studies in order to understand the applicability of superoscillatory signals.  This question can be reformulated in lay terms: by how much is a typical superoscillating signal worse than the optimal?
In section 2 we discuss the effect of noise on the distribution of the yield in the subspace of interpolating signals. We consider two common models of noise, namely a bimodal and a Gaussian distribution, and show that small deviations do not affect the yield, while larger deviations result in a yield which is orders of magnitude smaller, with a narrow transition zone between these two regimes. 
In section 3, we consider deviations from the Fourier components of the optimized superoscillatory signal in real space. Such deviations affect the quality of the interpolation, and thus may damage the superoscillatory nature of the signal. We therefore define and study in section 3.1 the interpolation sensitivity of the signal, which quantifies the extent to which the signal deviates from the interpolation constraints. We then explore in section 3.2 the impact of deviations from the Fourier components of the optimized signal on the superoscillatory yield. An interesting effect is that a moderate degree of noise can actually increase the yield, at the expense of modifying the shape, which could be of interest in applications. The last question we address in section 3.3 is the effect of these deviations on the number of oscillations that appear in the superoscillatory region. It turns out that large enough deviations can sometimes lower the number of oscillations rendering these signals not superoscillatory anymore.
In section 4, we summarize the paper and draw conclusions.

\section{Noise in the subspace of interpolating signals}

In order to progress, we consider the functional dependence of the yield on the constrained and on the unconstrained $B$'s. This composition is useful because any choice of the unconstrained $B$'s yields a signal which obeys the constraints.

The numerator $I$ on the right hand side of equation (\ref{eq:6}), can be expressed as follows
\begin{equation}
I=\sum\limits_{m.n=0}^{N} \Delta_{mn} A_m A_n=\sum\limits_{m,n=0}^{N}\Delta^{(R)}_{mn} B_m B_n \, ,
\label{eq:9}
\end{equation}
where ${{\mathbf{\Delta}}^{(\mathbf{R})}}=\mathbf{R\Delta}{\mathbf{R}}^{-1}$ , $\mathbf{R}$ being the rotation that takes the coefficient vector $\mathbf{A}$ into $\mathbf{B}$. Let us describe the matrix ${{\mathbf{\Delta}}^{(\mathbf{R})}}$ by the following block form
\begin{equation}
{{\bf{\Delta }}^{({\bf{R}})}} = \left( \begin{array}{l}
 {{\bf{\tilde \Delta}}_{(N + 1 - M) \times (N + 1 - M)}} \\ 
 {\bf{\Gamma}}^T_{M \times (N + 1 - M)} \\ 
 \end{array} \right.\left. \begin{array}{l}
 {{\bf{\Gamma }}_{(N + 1 - M) \times M}} \\ 
 {{\bf{\bar \Delta}}_{M \times M}} \\ 
 \end{array} \right) \, ,
\label{eq:10}
\end{equation}
where ${\bf{\Gamma}}^T$ is the transpose of ${\bf {\Gamma}}$. The yield can be now expressed in terms of the unconstrained $B'$s as
\begin{equation}
Y = \frac{I}{D}={\textstyle{{\sum\limits_{m,n = 0}^{N - M} {{{\tilde \Delta }_{mn}}{B_m}{B_n} + 2\sum\limits_{m = 0}^{N - M} {\sum\limits_{\scriptstyle n =  \hfill \atop \scriptstyle N + 1 - M \hfill}^N {{\Gamma _{mn}}{B_n^C}{B_m} + \sum\limits_{\scriptstyle m,n =  \hfill \atop \scriptstyle N + 1 - M \hfill}^N {{{\bar \Delta }_{mn}}{B_m^C}{B_n^C}} } } } } \over {\sum\limits_{m = 0}^{N - M} {B_m^2 + \sum\limits_{m = N + 1 - M}^N {(B_m^C)^2} } }}} \, .
\label{eq:11}
\end{equation}
To simplify the notation we rewrite
\begin{equation}
Y = {\textstyle{{\sum\limits_{m,n = 0}^{N - M} {{{\tilde \Delta}_{mn}}{B_m}{B_n}}} + \sum\limits_{m = 0}^{N - M} {\alpha_m}{B_m} + D_1} \over {\sum\limits_{m = 0}^{N - M} {B_m^2 + D_2}}} \, .
\label{eq:12}
\end{equation}
where the symmetric positive definite matrix $\tilde \Delta $, the coefficients $\alpha_m$ and the constants $D_1$ and $D_2$ depend on the interpolation constraints (and also on the specific frame of reference used in the $S_M$ subspace). 

We would like to take a random set of $B$'s and get an idea of the corresponding random yield to which they give rise. Such randomness serves two purposes. First, in many physical systems noise is prevalent and it is of great importance to understand its impact on the superoscillatory signal and in particular on its yield. Second, it provides some insight into the orders of magnitude of $B$'s which are anticipated in the context of interpolating signals.

It can be shown that $\alpha_m$, $D_1$ and $D_2$ are typically much smaller than the quadratic terms in Eq. (\ref{eq:12}). This means that for a random choice of the $B$'s the yield would be typically given by
\begin{equation}
Y \approx  {\textstyle{{\sum\limits_{m,n = 0}^{N - M} {{{\tilde \Delta}_{mn}}{B_m}{B_n}}}} \over {\sum\limits_{m = 0}^{N - M} {B_m^2}}} \, .
\label{eq:13}
\end{equation}
Thus, it is clear that typically, the value of $Y$ is distributed between the largest and smallest eigenvalue of $\tilde \Delta $, which spans many decades (the condition number of $\tilde \Delta$, defined as the ratio between the largest and smallest eigenvalue is very large, and so the problem is typically ill-conditioned). To give just two examples, the largest and smallest eigenvalues of $\tilde \Delta $ $\left( \delta_{\max},\delta _{\min} \right)$ for the cases of $N = 6,M = 4$  and $N = 9,M = 6$ are given by $\left( {0.000179,3.12{\rm{ }}{{10}^{ - 14}}} \right)$ and $\left( {4.65{\rm{ }}{{10}^{ - 6}},3.64{\rm{ }}{{10}^{ - 21}}} \right)$ respectively. It is interesting to note that the corresponding optimal values of the yield are $Y = 0.0138$ (for $N = 6,M = 4$) and $Y = 0.0004$ (for $N = 9,M = 6$). This implies that a random choice of a signal that interpolates the constrained points gives a yield which is {\bf orders of magnitude} less than the optimal one, which emphasizes the necessity to optimize the superoscillatory signals.

At this point we can say more about the statistical properties of the yield of interpolating signals. We will assume first that the $B$'s are independent and drawn from a distribution
\begin{equation}
P(B) = \frac{1}{\sigma }g\left( {\frac{B}{\sigma }} \right) \, ,
\label{eq:14}
\end{equation}
where $g(x)$ is a normalized and symmetric distribution, with variance equal to $1$. Therefore, $\sigma$ measures the width of $P(B)$. Such noise describe possible situations where the optimized set of $B$'s are corrupted to a certain extent, or that their accuracy has been degraded. Independence of the $B$'s simply assumes that the a random bias has been introduced directly in the subspace of interpolating signals. Namely, such noise will not modify {\it at all} the fact that the signal generated by this set always obeys the superoscillatory constraints.

We will consider averages as a function of $\sigma $ and then distributions for fixed $\sigma $. We start with some obvious results: it is clear that
\begin{equation}
\mathop {\lim }\limits_{\sigma  \to 0} Y = \frac{{{D_1}}}{{{D_2}}} \, .
\label{eq:15}
\end{equation}
This gives, combined with equation (\ref{eq:13}), a lower bound on the optimal yield, namely 
\begin{equation}
Y_{opt} > \max \left( {\frac{{{D_1}}}{{{D_2}}},{\delta _{\max }}} \right) \, .
\label{eq:16}
\end{equation}

\noindent
Consider next the very simple average of the total energy per period (the denominator of $Y$ – eq.~(\ref{eq:5})) 
\begin{equation}
\left\langle E \right\rangle  = (N + 1 - M){\sigma ^2} + D_2 \, ,
\label{eq:17}
\end{equation}
which turns out to be independent of the specific distribution. The average yield is however more difficult and does depend on the distribution. Let us therefore discuss a two illustrative examples, namely a bi-modal distribution, which is a discrete distribution, and a Gaussian distribution, which is continuous. These two distributions represent a range of possible symmetric distributions with a finite variance, which are often used to model noise.

Consider first the bimodal distribution, 
\begin{equation}
P(B) = \frac{1}{{2\sigma }}\left[ {\delta \left( {\frac{B}{\sigma } - 1} \right) + \delta \left( {\frac{B}{\sigma } + 1} \right)} \right] ,
\end{equation}

\noindent
where $\delta(x)$ is the Dirac delta function. The average yield in this case is simply
\begin{equation}
\left\langle Y \right\rangle  = \frac{{\left( {tr\tilde \Delta } \right){\sigma ^2} + {D_1}}}{{(N + 1 - M){\sigma ^2} + D_2}} \, .
\label{eq:18}
\end{equation}

\noindent
This is nothing but a Sigmoid function, attaining the value $D_1/D_2$ when $\sigma \rightarrow 0$ (in agreement with Eq. (\ref{eq:15}), and $tr\tilde \Delta /(N+1-M)$ for $\sigma \rightarrow \infty$. Actually $tr\tilde \Delta$ is very small for superoscillatory signals \cite{Katzav13}, and thus $\left\langle Y \right\rangle$ is monotonically decreasing, interpolating between $D_1/D_2$ and $tr\tilde \Delta /(N+1-M) \ll 1$.

\noindent
Consider next a Gaussian distribution, 
\begin{equation}
P(B) = \frac{1}{{\sqrt {2\pi \sigma } }}\exp \left( { - \frac{{{B^2}}}{{2\sigma^2 }}} \right).
\end{equation}
Although considerably more complicated, the average yield can also be obtained analytically here,
\begin{equation}
\left\langle Y \right\rangle  = \frac{{{D_1}}}{{2{\sigma ^2}}}{e^{\frac{{{D_2}}}{{2{\sigma ^2}}}}}{E_{\frac{{N + 1 - M}}{2}}}\left( {\frac{{{D_2}}}{{2{\sigma ^2}}}} \right) + \frac{{tr\tilde \Delta }}{2}{e^{\frac{{{D_2}}}{{2{\sigma ^2}}}}}{E_{\frac{{N + 3 - M}}{2}}}\left( {\frac{{{D_2}}}{{2{\sigma ^2}}}} \right)  \, .
\label{eq:19}
\end{equation}
where ${E_\nu }\left( z \right) = \int\limits_1^\infty  {dx\frac{\exp (-xz)}{x^\nu}} $, is the exponential integral function \cite{NIST}. In order to obtain more insight, we can study the various limits o this expression. In fact, $E_\nu(z) \simeq e^{-z}/z$ for $z \rightarrow \infty$ and therefore, it is easy to verify that here too $\langle Y \rangle \rightarrow D_1/D_2$ in the limit $\sigma \rightarrow 0$ (consistent with Eq. (\ref{eq:15})). In the other limit, namely $\sigma \rightarrow \infty$, we can use $E_\nu(z) \simeq 1/(\nu-1)$ for $z \rightarrow 0$ (and $\nu \ge 1$). Since $N \ge M+1$, this gives rise to $\langle Y \rangle \rightarrow  \tr\tilde \Delta /(N+1-M)$. Also, by differentiating Eq. (\ref{eq:19}) with respect to $\sigma$ it is easy to show that it is a monotonically decreasing function, since $dE_\nu(z)/dz = -E_{\nu-1}(z)$ which is negative definite for all real values of $z$ .

\begin{figure}[!t]
\centering
\centerline{\includegraphics[width=4in]{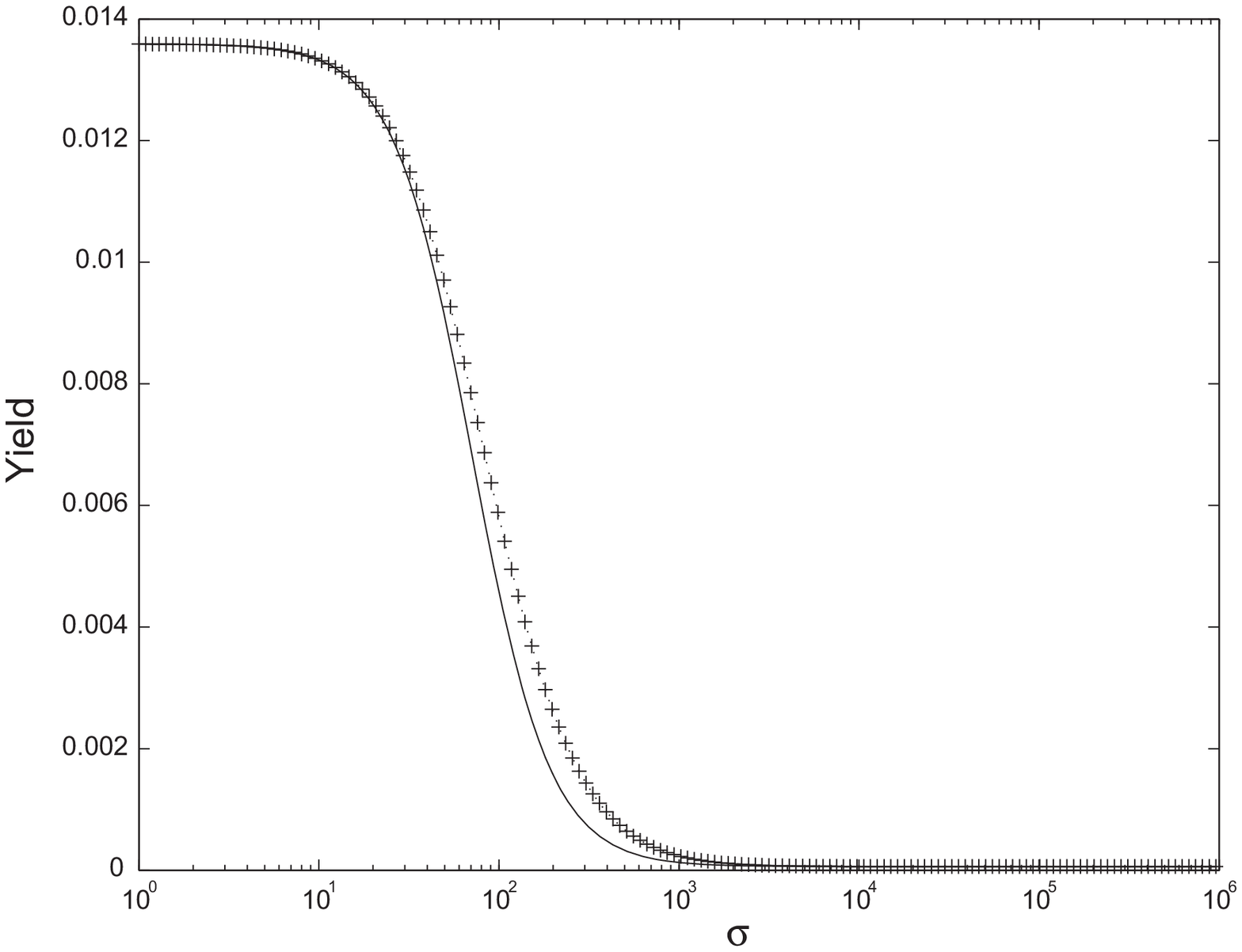}}
\caption{The average yield $\left\langle Y \right\rangle $ as a function of $\sigma $ for $N = 6,M = 4$. The upper curve is for the Gaussian distribution (given by eq.~(\ref{eq:19})) and the lower one for the bimodal distribution (given by eq.~(\ref{eq:18})).}
\label{fig:1}
\end{figure}

\begin{figure}[!t]
\centering
\centerline{\includegraphics[width=4in]{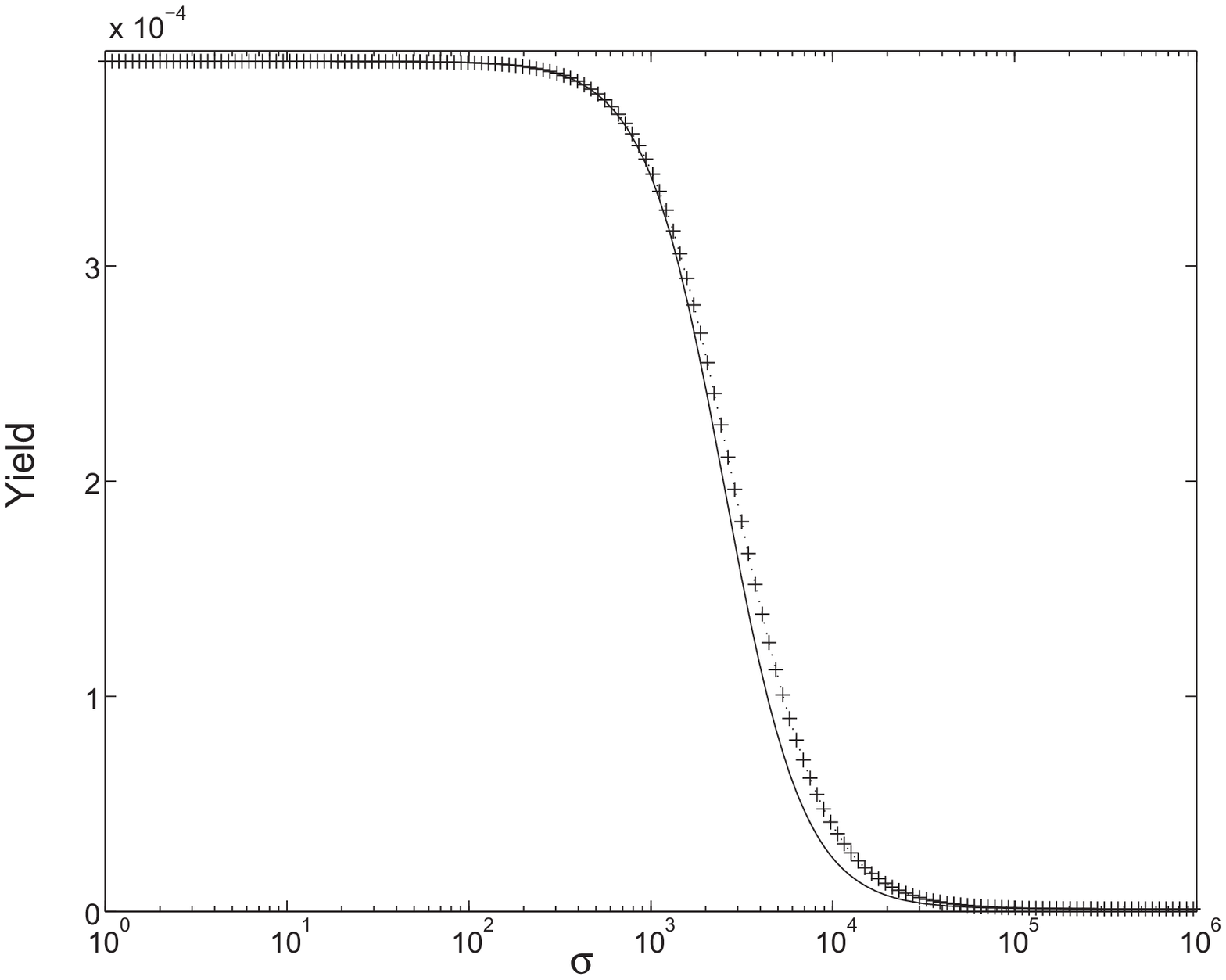}}
\caption{The average yield $\left\langle Y \right\rangle $ as a function of $\sigma $ for $N = 9,M = 6$. The upper curve is for the Gaussian distribution (given by eq.~(\ref{eq:19})) and the lower one for the bimodal distribution (given by eq.~(\ref{eq:18})).}
\label{fig:2}
\end{figure}

It is clear from equations (\ref{eq:13}) and (\ref{eq:15}) that regardless of the distribution the average has to interpolate between $D_1/D_2$ at small $\sigma $ and $tr\tilde \Delta/(N + 1 - M)$ (which is known to be a very small number \cite{Katzav13}), and are monotonically decreasing. The two averages, (\ref{eq:18}) and (\ref{eq:19}) are presented in Figs. \ref{fig:1} and \ref{fig:2} for the cases $N = 6,M = 4$ and $N = 9,M = 6$, respectively. 
The figures suggest that although $\left\langle Y \right\rangle $ depends on the precise distribution of the $B$'s, this dependence is not very pronounced. Both cases result in $\langle Y \rangle $ which behaves like a sigmoid, namely it is stays very close to $\langle Y \rangle (\sigma  = 0)=D_1/D_2$ until it drops to zero over a relatively narrow range. This is indicative of the presence of a sharp crossover as a function of the noise level.

\begin{figure}[!t]
\centering
\centerline{\includegraphics[width=4in]{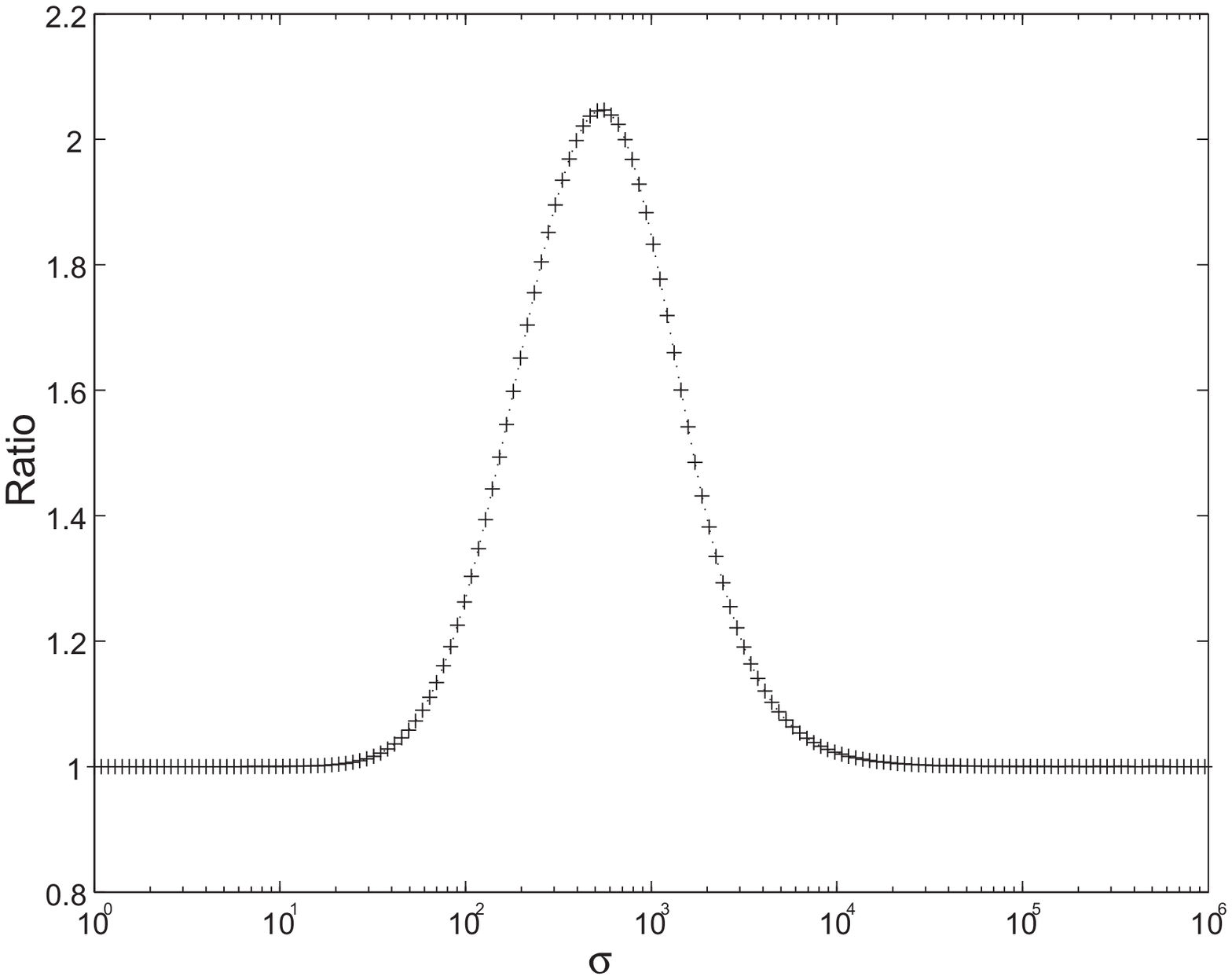}}
\caption{The ratio between the average yield for the Gaussian distribution (given by eq.~(\ref{eq:19})) and that for the bimodal distribution (given by eq.~(\ref{eq:18})) as a function of $\sigma $ for $N = 6,M = 4$.}
\label{fig:3}
\end{figure}

The two figures are not conclusive concerning whether for large $\sigma $ the ratio of the average yield obtained from the Gaussian distribution to that obtained from the bimodal distribution tends to $1$, as predicted above (Figs. \ref{fig:1} and \ref{fig:2} only show that both tend to zero). Therefore, we present the ratio of the two for the case $N = 6,M = 4$ in Fig. \ref{fig:3} (the same figure for the case $N = 9,M = 6$ is quite similar and will not be presented here for brevity).

\begin{figure}[!t]
\centering
\centerline{\includegraphics[width=4in]{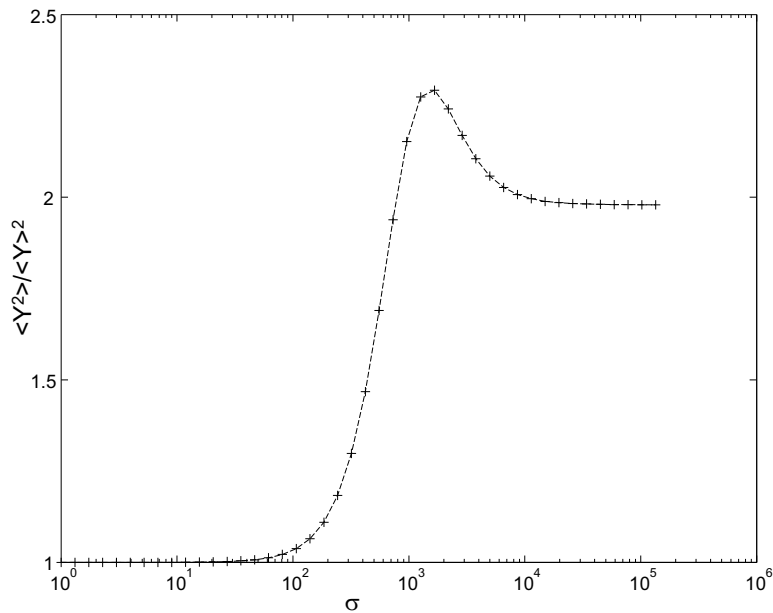}}
\caption{The ratio between the second moment of the yield and the square of the mean $\left\langle Y^2\right\rangle /  \left\langle Y \right\rangle ^2$ for the bimodal distribution as a function of $\sigma $ for $N = 6,M = 4$.}
\label{fig:4}
\end{figure}

\begin{figure}[!t]
\centering
\centerline{\includegraphics[width=4in]{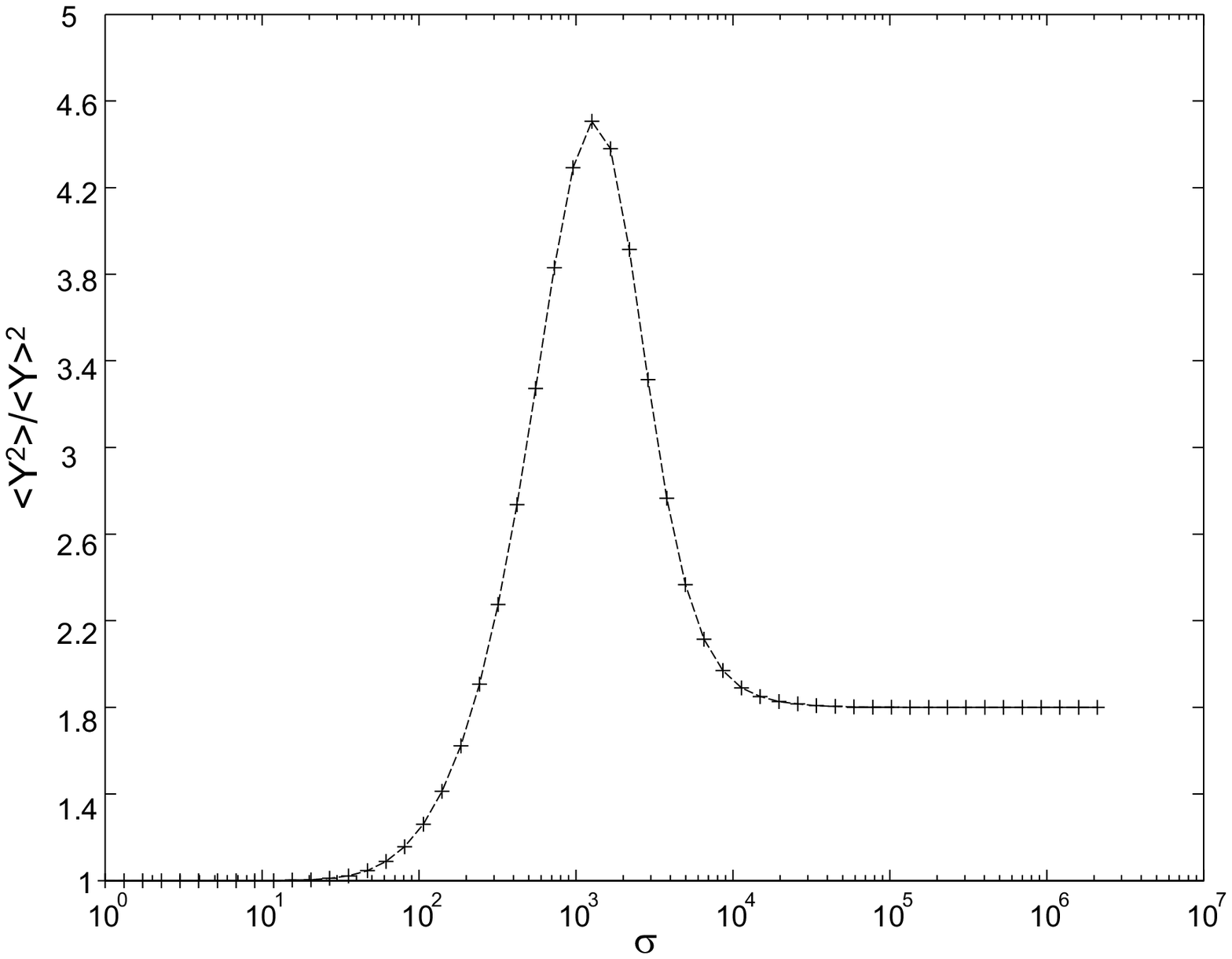}}
\caption{The ratio between the second moment of the yield and the square of the mean $\left\langle Y^2\right\rangle /  \left\langle Y \right\rangle ^2$ for the Gaussian distribution as a function of $\sigma $ for $N = 6,M = 4$.}
\label{fig:5}
\end{figure}

\begin{figure}[!t]
\centering
\centerline{\includegraphics[width=2.5in]{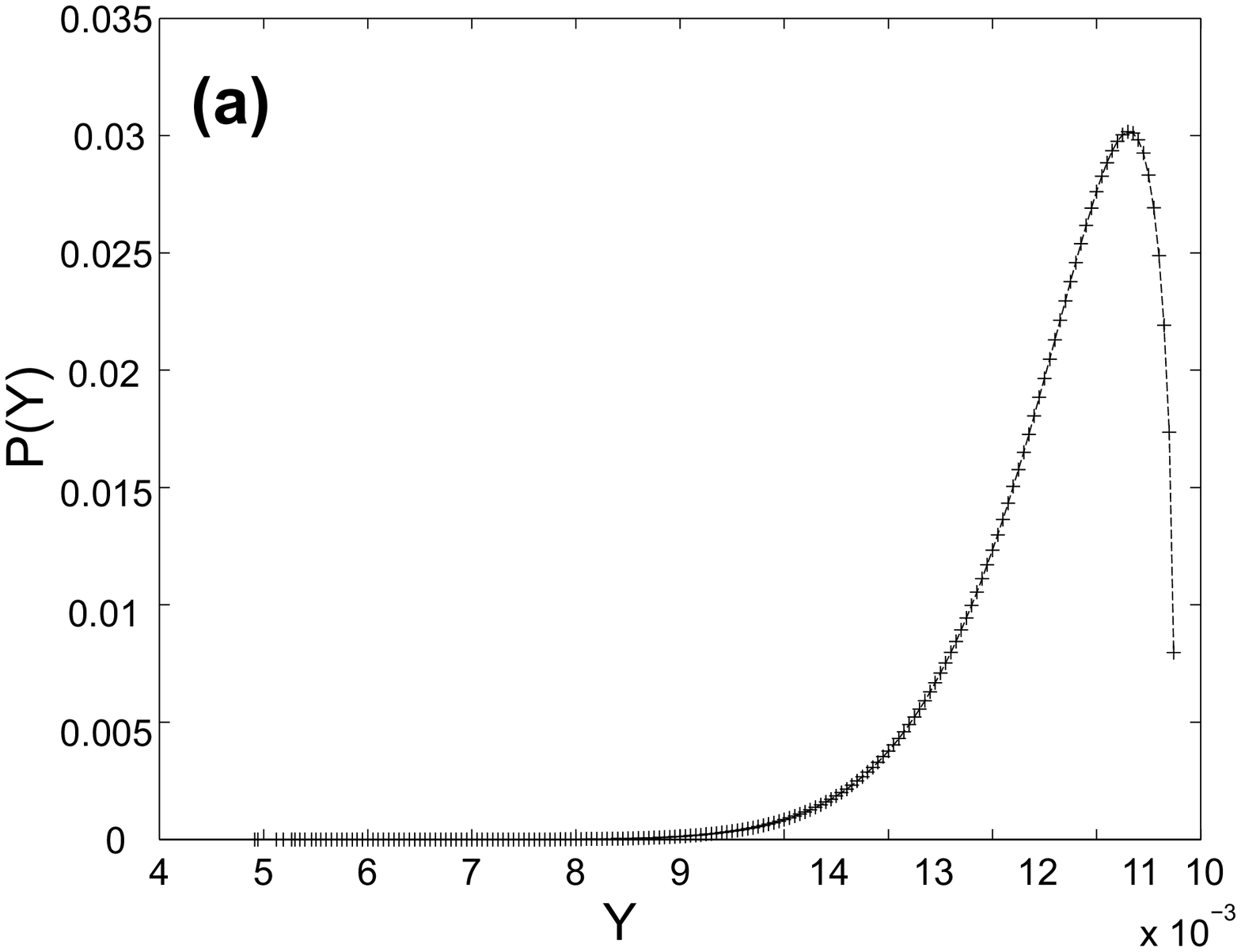}\includegraphics[width=2.5in]{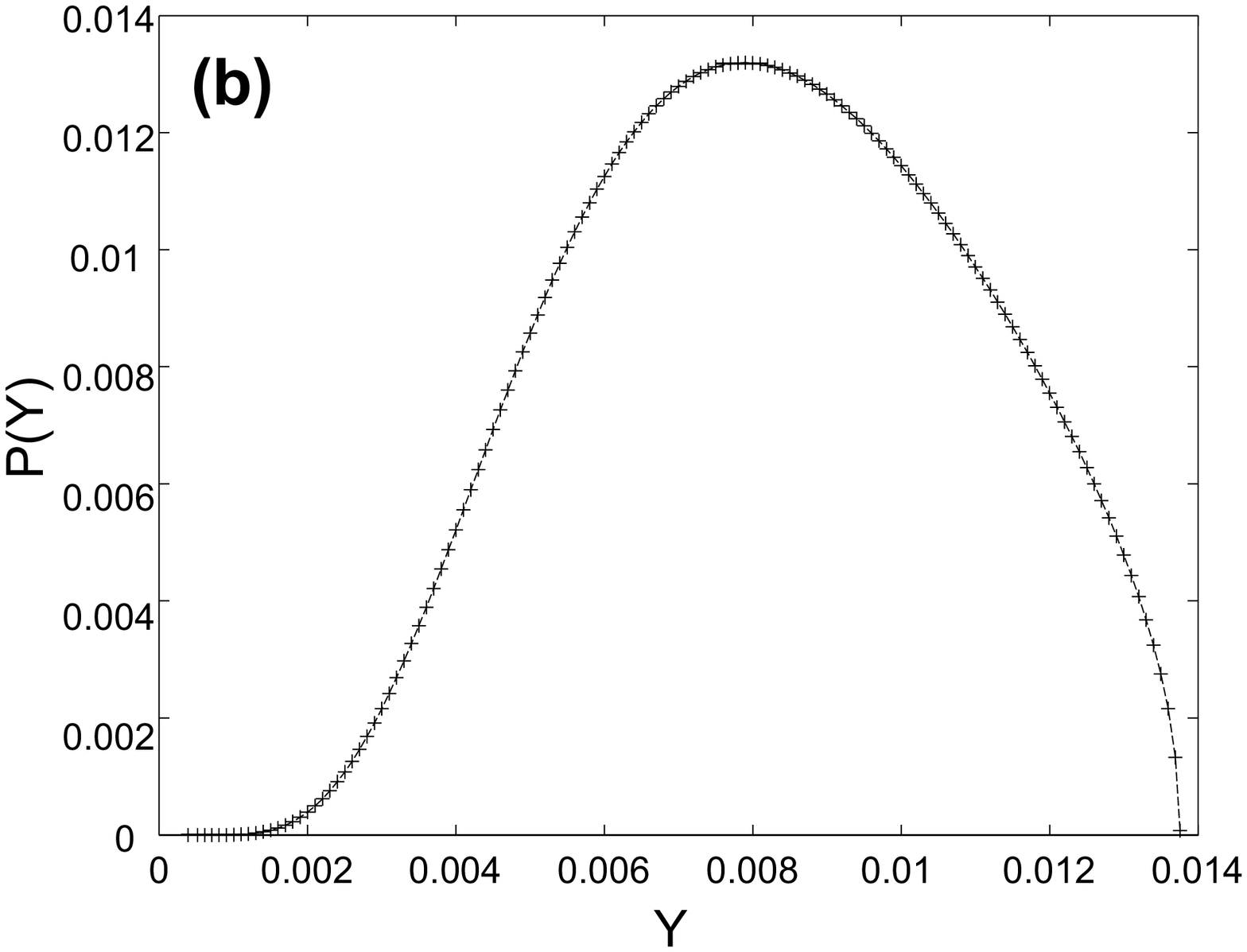}\includegraphics[width=2.5in]{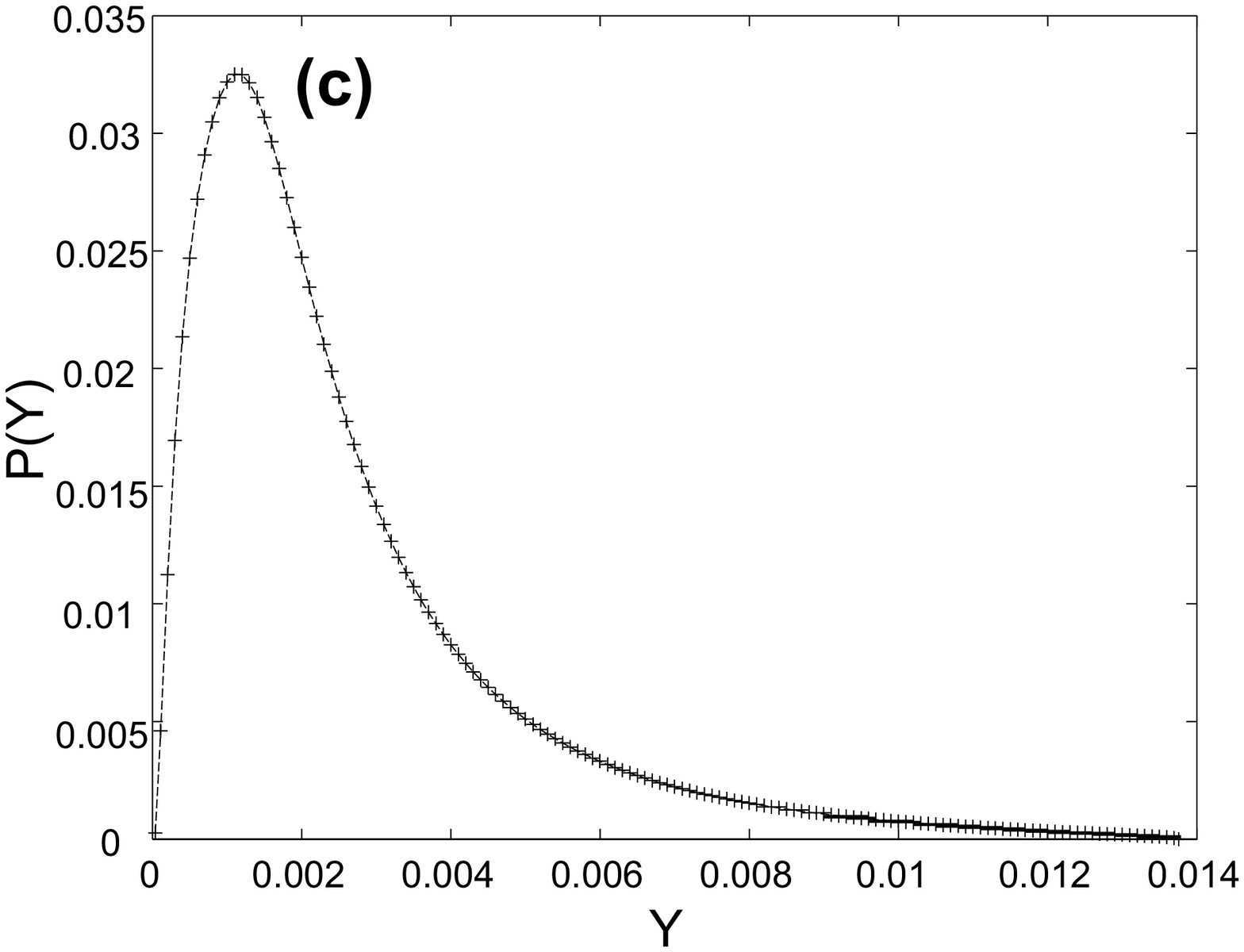}}
\caption{The distribution of the Yield for three different representative values of $\sigma $ for $N = 6,M = 4$: (a) $\sigma  = 20$, (b) $\sigma  = 65$ and (c) $\sigma  = 200$, corresponding to the three regions identified in Fig. \ref{fig:5}. As can be seen, the distributions are indeed very different with pronounced skewness in opposite directions.}
\label{fig:6}
\end{figure}

It is of interest to obtain also the second moment of the yield. This calculation can also be performed analytically even for the Gaussian distribution but we will not present it in the following, because it is considerably more complicated and lengthy. In Figs. \ref{fig:4} - \ref{fig:5} we present the ratio of second moment to the square of the first moment (namely, $\left\langle Y^2 \right\rangle  \mathord{\left/ {\vphantom {{\left\langle {{Y^2}} \right\rangle } {\left\langle Y \right\rangle ^2}}} \right. \kern-\nulldelimiterspace} {\left\langle Y \right\rangle ^2}$) for the case $N = 6,M = 4$ and the two distributions.

Considering Figs. \ref{fig:4} - \ref{fig:5} it becomes clear that the yield distribution behaves differently for three distinct regions of $\sigma$. Indeed, that is the case as presented in Fig. \ref{fig:6} depicting yield distributions for three representative values of $\sigma$.

\begin{figure}[!t]
\centering
\centerline{\includegraphics[width=4in]{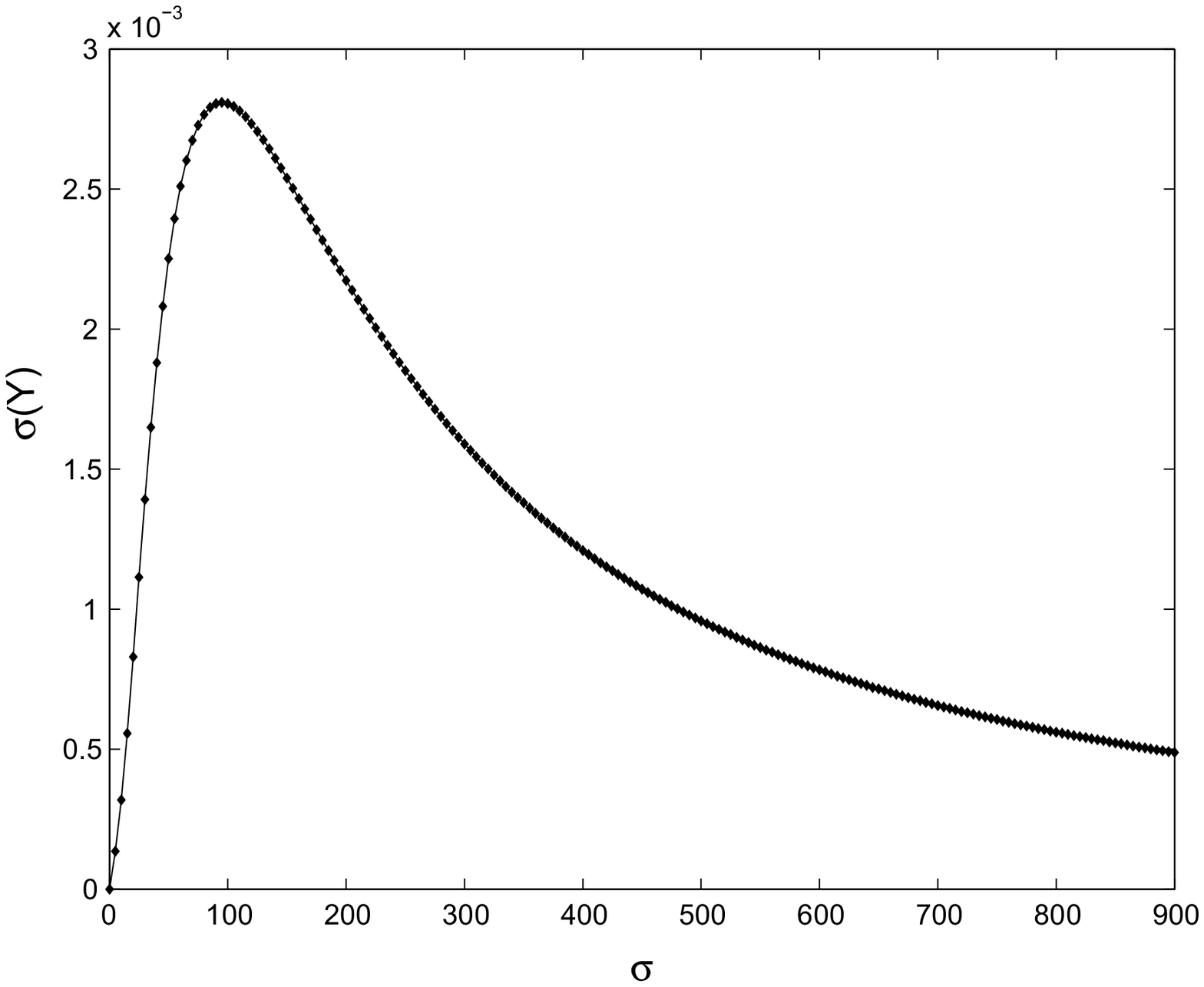}}
\caption{The standard deviation of the yield $\sigma(
Y) = \sqrt{\left \langle Y^2 \right \rangle - \left \langle Y \right \rangle^2}$ as a function of $\sigma $ for $N = 6,M = 4$. As can be seen this is a monotonously decreasing function.}
\label{fig:61}
\end{figure}

It follows from the above discussion that the average yield is a monotonically decreasing functions of $\sigma$. It is actually also the case for its standard deviation $\sigma(Y) = \sqrt{\left \langle Y^2 \right \rangle - \left \langle Y \right \rangle^2}$ as demonstrated in Fig. \ref{fig:61} for $N=6$, $M=4$, and can be proven more generally for the Bimodal and the Gaussian cases. 

At this point, a natural question that arises is whether the optimal yield is characterized by very small $B$'s? 

\noindent
To answer this question, consider the total energy (per period)
\begin{equation}
E = D = \sum\limits_{m = 0}^{N - M} {B_m^2 + D_2} \, .
\label{eq:20}
\end{equation}
It is clear that the total energy is minimized when ${B_m} = 0$ for $0\le m \le N - M$ and the minimal total energy per period is $D_2$. We note further that the total energy can be expressed as $E = \frac{I}{Y}$. Thus the (very simple) minimization process of $E$ with respect to the choice of the $B_m$'s , can be expressed as 
\begin{equation}
\frac{1}{Y}\frac{{\partial I}}{{\partial {B_m}}} - \frac{I}{{{Y^2}}}\frac{{\partial Y}}{{\partial {B_m}}} = 0 \, .
\label{eq:21}
\end{equation}
Consequently, on the set of $B$'s for which the total energy per period is minimized we have
\begin{equation}
\frac{{\partial Y}}{{\partial {B_m}}} = Y\frac{{\partial I}}{{\partial {B_m}}} \, ,
\label{eq:22}
\end{equation}
and since $Y$ is typically very small (in the superoscillatory regime), we conclude that at the point of minimal total energy, the yield is almost extremal (i.e., maximal) too, because the absolute value of each of its derivatives is very small. Conversely, the $B$'s at the optimal yield are very small, as expected from the $\sigma $ dependence of the average yield. Thus, optimization of the total energy under the constraints is typically almost equivalent to optimization of the yield itself.

\section{Noise in the Fourier coefficients of the signal}
So far, we have considered the behaviour of the yield in the space of interpolating signals, i.e. in the subspace where any choice of the $B$'s yields a signal which automatically obey the constraints. Next we address the question of possible departure of the signal from the prescribed interpolation, in the vicinity of the yield optimized interpolating signal. Such departure will no doubt happen in any physical device that would need to utilize such signals, because any realization of a superoscillatory signal using a Fourier representation (\ref{eq:1}) would have access to the set of $A$'s with finite precision (due to the presence of noise for example). Such limited precision will obviously affect the yield, but of not less importance, it could also imply violation of the interpolation constraint, which may eventually destroy the superoscillatory nature of the signal. Recall that the yield is maximized with respect to the choice of the ${A_m}$'s under the constraints given by equations (\ref{eq:2})-(\ref{eq:3}). The result is a set of optimal Fourier coefficients $\left\{ {{{\tilde A}_m}} \right\}$. To address the questions of violation of the constraints due to finite precision, assume that the actual Fourier coefficients describing the superoscillating function are given by
\begin{equation}
{\hat A_m} = {\tilde A_m} + {\delta _m} \, ,
\label{eq:23}
\end{equation}
where ${\delta _m}$'s are random errors and governed by a Gaussian distribution with standard deviation $\delta $.

\subsection{The interpolation sensitivity of the signal}
Consider the first question, which is the sensitivity of the interpolation to random fluctuations around the optimal solution. Equations (\ref{eq:2})-(\ref{eq:3}) imply that the change in the value of the function $f(t)$ at the points ${t_j} = ja/(M-1)$ for $j = 0,...,M - 1$, is given by 
\begin{equation}
\delta {f_j} = \sum\limits_{m = 1}^N {{C_{jm}}{\delta _m}} \, ,
\label{eq:24}
\end{equation}
where the absolute values of all the ${C_{jm}}$'s are less or equal to $1$ (by definition -- see eq.~(\ref{eq:2})). Consequently, 
\begin{equation}
\left| {\delta {f_j}} \right| \propto \delta \sqrt N \, .
\label{eq:25}
\end{equation}

\noindent
As long as the right hand side is (considerably) less than one we still have
\begin{equation}
sign[f({t_j})] = {( - 1)^j} \, ,
\label{eq:26}
\end{equation}
so that the number of oscillations in the interval $[0,a]$ will not decrease and thus the required superoscillation is still preserved.

To what extent do the fluctuations around the yield-optimized signal affect the required superoscillation? Put differently, what is the error level $\delta$ that would start impairing the required superoscillations? From eqs.~(\ref{eq:25})-(\ref{eq:26}) it is clear that the required absolute accuracy is
\begin{equation}
\delta  < 1/\sqrt N \, ,
\label{eq:27}
\end{equation}
which does not seem to pose a serious problem.

Note, however, that the problem might be that the error $\delta $ can be reduced only below a value which is relative to the typical absolute value of the $A$'s. If that is the case it might pose a severe problem, because the optimal Fourier amplitudes are exponentially large in $N$ (for superoscillating signals) and this is why out of the superoscillating interval the absolute value of $f$ is exponentially large in $N$. In that case, the relative accuracy required in the Fourier coefficients is of the order of $\theta  = \delta /\left| A \right| \propto {N^{-1/2}}{\alpha^{-N}}$, where $\left| A \right|$ denotes the typical size of the ${\tilde A_m}$'s and $\alpha  > 1$. Thus, if it is only relative accuracy that is possible to control, care should be taken that $N$ is not too large in order to preserve the superoscillations. If absolute accuracy can be obtained, regardless of the magnitude of the optimal Fourier coefficients, it should be no problem to preserve superoscillations under very reasonable fluctuations in those quantities. We characterize the departure from the enforced interpolation, for a given set of errors in the Fourier coefficients by a single parameter, $\delta V$, which we call the interpolation sensitivity, and is defined as
\begin{equation}
\delta V = \frac{1}{M}\sum\limits_{i = 0}^{M - 1} {{{\left( {\delta {f_i}} \right)}^2}} \, .
\label{eq:28}
\end{equation}

\begin{figure}[!t]
\centering
\centerline{\includegraphics[width=4in]{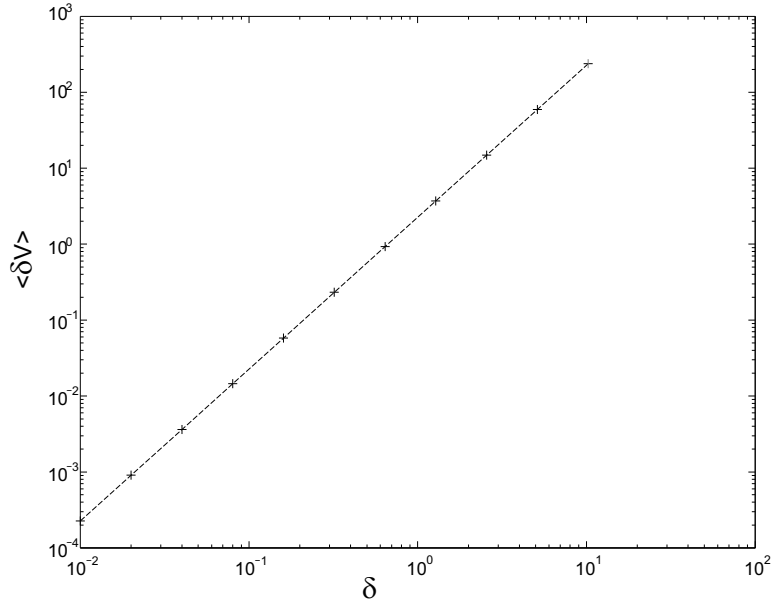}}
\caption{The mean interpolation sensitivity $\left\langle {\delta V} \right\rangle $ as a function of $\delta $ for $N = 6, M = 4$.}
\label{fig:7}
\end{figure}

\begin{figure}[!t]
\centering
\centerline{\includegraphics[width=4in]{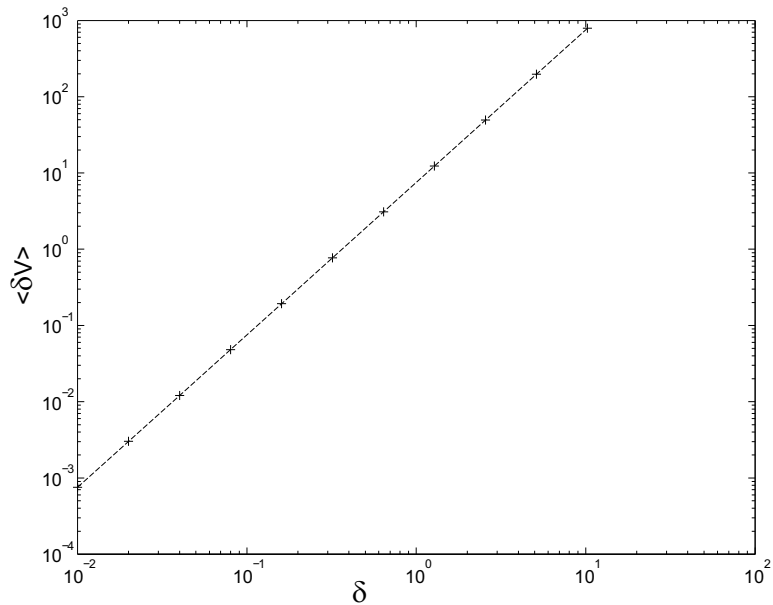}}
\caption{The mean interpolation sensitivity $\left\langle {\delta V} \right\rangle $ as a function of $\delta $ for $N = 12, M = 8$}
\label{fig:8}
\end{figure}

\noindent
Figures \ref{fig:7} - \ref{fig:8} show the interpolation sensitivity, which is the average of $\delta V$ over the Gaussian distribution of the ${\delta _m}$'s, as a function of $\delta $ for two examples of yield optimized superoscillations, 

Note that in spite of the fact that in the graphs above $\delta$ attains values up to $10^2$, only for $\delta$'s considerably smaller than one, we may expect the original frequency in the superoscillating interval to be unchanged. (The oscillation period, $T$, is taken as the average peak to peak distance and the frequency is $\omega  = 2\pi /T$.)

While the interpolation sensitivity $\left\langle {\delta V} \right\rangle $ is an important quantity that characterizes the deviations from the enforced interpolation, it is, obviously, not less important to obtain the distribution of the quantity $\delta V$. In the following we give this distribution for the case $N = 12,{\rm{ }}M = 8$ and for a specific value of $\delta  = 0.01$. It is clear that at least as far as the order of magnitude is considered, the interpolation sensitivity represents well the distribution of the values of $\delta V$. The distribution is obtained by considering $10^7$ realizations of the set of the low frequency Fourier coefficients.
\begin{figure}[!t]
\centering
\centerline{\includegraphics[width=4in]{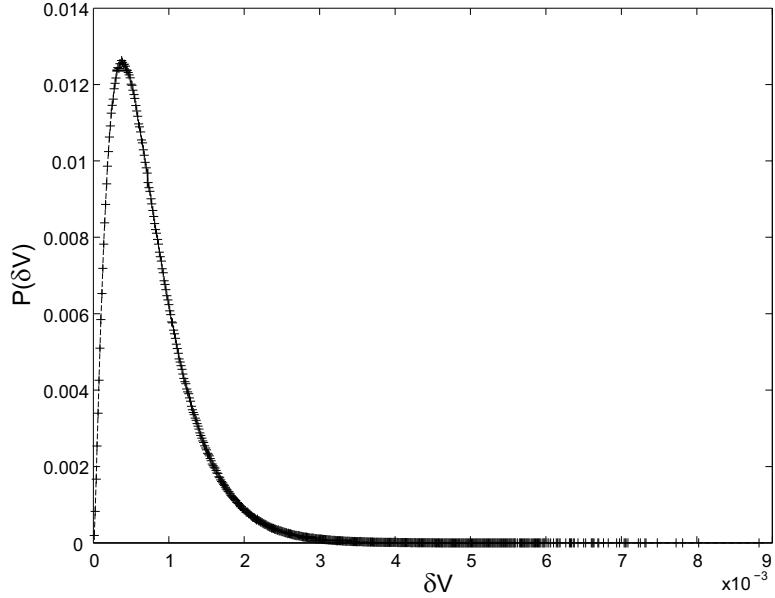}}
\caption{The normalized distribution, $P\left( {\delta V} \right)$, for $\delta  = 0.01$. The distribution is obtained from ${10^7}$ realizations of the set of Fourier coefficients.}
\label{fig:9}
\end{figure}

\subsection{Sensitivity of the Yield}
\begin{figure}[!t]
\centering
\centerline{\includegraphics[width=4in]{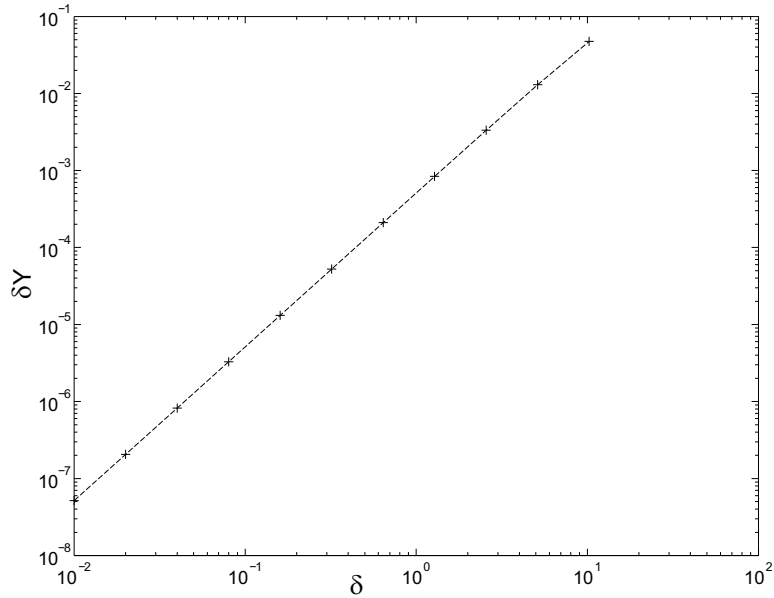}}
\caption{The average deviation of the yield as a function of $\delta $ for $N = 6, M = 4$.}
\label{fig:10}
\end{figure}

\begin{figure}[!t]
\centering
\centerline{\includegraphics[width=4in]{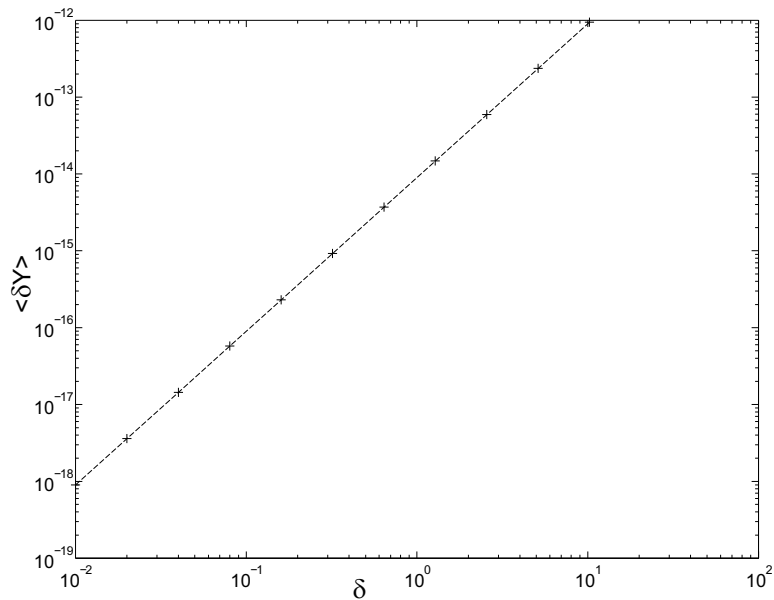}}
\caption{The average deviation of the yield as a function of $\delta $ for $N = 12, M = 8$.}
\label{fig:11}
\end{figure}

The next problem is that of the sensitivity of the superoscillation yield. The yield can be expressed in terms of the deviations as follows. The deviations from the optimal Fourier coefficients define a function giving the deviation from the optimal function, $\tilde f(t)$, obeying the constraints, namely
\begin{equation}
\delta (t) = \frac{{{\delta _0}}}{{\sqrt {2\pi } }} + \sum\limits_{m = 1}^N {\frac{{{\delta _m}}}{{\sqrt \pi  }}} \cos (mt) \, .
\label{eq:29}
\end{equation}
The first order change in the yield can be written in terms of this function as
\begin{equation}
\delta Y = 2\frac{{\int\limits_{ - a}^a {\tilde f(t)\delta (t)dt} }}{{\int\limits_{ - \pi }^\pi  {{{\tilde f}^2}(t)dt} }} - 2Y\frac{{\int\limits_{ - \pi }^\pi  {\tilde f(t)\delta (t)dt} }}{{\int\limits_{ - \pi }^\pi  {{{\tilde f}^2}(t)dt} }} \, .
\label{eq:30}
\end{equation}
The fact that the yield has a linear correction when calculated at its maximum, should not be surprising, because it is a constrained maximum and the linear part appears as a result of the small violations of the strict interpolation constraint. Had we considered deviations in the Fourier coefficients that respect the constraints, the lowest order correction would be quadratic in $\delta (t)$. This implies that because of the departure from the interpolation constraints, the yield can increase under fluctuations. In Figs. \ref{fig:10} and \ref{fig:11} we present the average of $\delta Y$, as a function of $\delta $. As we shall see, the average, which is not the average of the linear approximation of $\delta Y$ given above, is positive. This will be explained later when the scatter graph, giving the distribution of pairs $\left( {\delta V,\delta Y} \right)$ for given $\delta $ will be discussed. 

The relative deviation in the yield is given by
\begin{equation}
\frac{{\delta Y}}{Y} = 2\frac{{\int\limits_{-a}^a {\tilde f(t)\delta (t)dt} }}{{\int\limits_{-a}^a {{{\tilde f}^2}(t)dt} }} - 2\frac{{\int\limits_{-\pi }^\pi  {\tilde f(t)\delta (t)dt} }}{{\int\limits_{-\pi }^\pi  {{{\tilde f}^2}(t)dt} }} \, .
\label{eq:31}
\end{equation}
An upper bound on the relative error in the yield can be obtained by noting first that a $t$-independent bound on the mean square deviation function is easily obtained from equation (\ref{eq:29}),
\begin{equation}
\left\langle {{\delta ^2}(t)} \right\rangle  < \frac{N}{\pi} \delta^2 \, .
\label{eq:32}
\end{equation}
Then the error in the yield clearly obeys
\begin{equation}
\frac{{\left| {\delta Y} \right|}}{Y} < 2\left[ {\frac{{\left| {\int\limits_{ - a}^a {\tilde f(t)\delta (t)dt} } \right|}}{{\int\limits_{ - a}^a {{{\tilde f}^2}(t)dt} }} + \frac{{\left| {\int\limits_{ - \pi }^\pi  {\tilde f(t)\delta (t)dt} } \right|}}{{\int\limits_{ - \pi }^\pi  {{{\tilde f}^2}(t)dt} }}} \right] \, ,
\label{eq:33}
\end{equation}
and by the Schwartz inequality and equation (\ref{eq:32}) above, 
\begin{equation}
\frac{{\left| \delta Y \right|}}{Y} < \frac{{4\sqrt N \delta }}{\sqrt \pi }\left[ {\sqrt {\frac{a}{{\int\limits_{ - a}^a {{{\tilde f}^2}(t)dt} }}}  + \sqrt {\frac{\pi }{{\int\limits_{ - \pi }^\pi  {{{\tilde f}^2}(t)dt} }}} } \right] \, .
\label{eq:34}
\end{equation}
The second term in the square brackets above can be neglected compared to the first and taking into account that within the superoscillating interval, ${\tilde f^2}(t)$ is of order $1$, we arrive at the conclusion that 
\begin{equation}
\frac{{\left| {\delta Y} \right|}}{Y} < \beta \sqrt {\frac{N}{a}} \delta \, ,
\label{eq:35}
\end{equation}
where $\beta$ is a constant of order $1$.

\begin{figure}[!t]
\centering
\centerline{\includegraphics[width=4in]{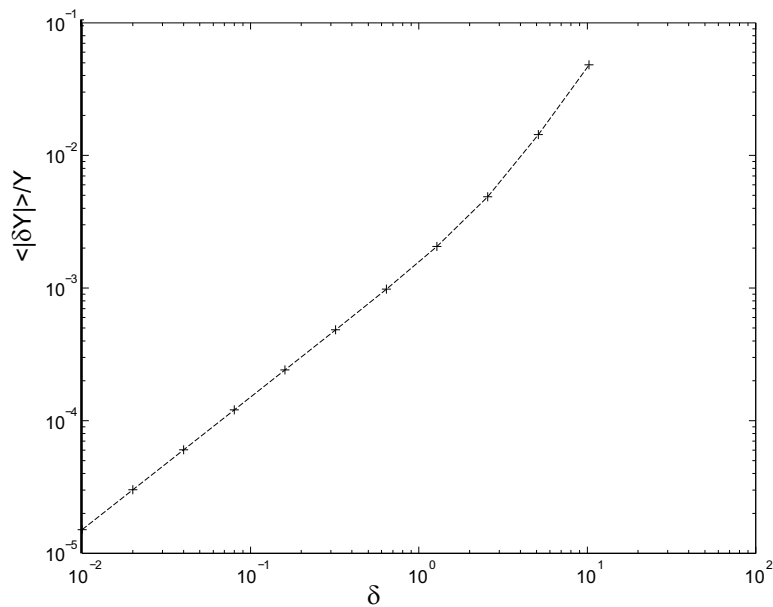}}
\caption{The yield sensitivity $\left\langle |\delta Y| \right\rangle  / Y$ for $N = 6, M = 4$.}
\label{fig:12}
\end{figure}

\begin{figure}[!t]
\centering
\centerline{\includegraphics[width=4in]{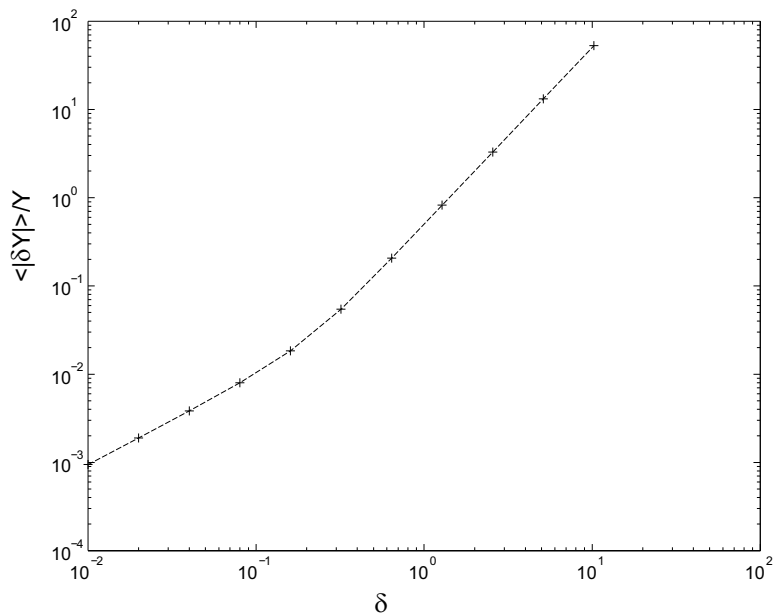}}
\caption{The yield sensitivity $\left\langle |\delta Y| \right\rangle  / Y$ for $N = 12, M = 8$.}
\label{fig:13}
\end{figure}

Thus, if the condition for the preservation of superoscillation in the presence of additive errors in the optimal Fourier coefficients $\left\{ {{{\tilde A}_m}} \right\}$ is obeyed, it ensures automatically a good upper bound on the error in the yield. In Figs. \ref{fig:12} - \ref{fig:13} we show the average relative error in the yield, which we call the yield sensitivity.

\begin{figure}[!t]
\centering
\centerline{\includegraphics[width=4in]{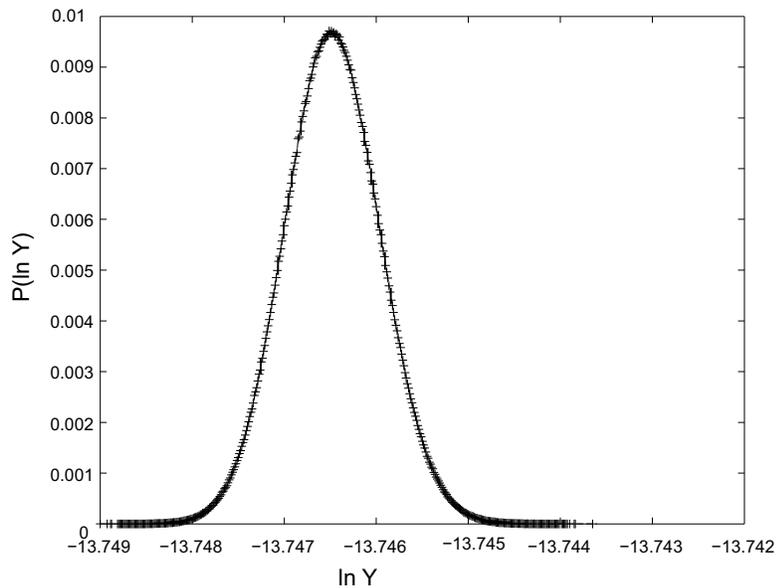}}
\caption{The normalized distribution of the logarithm of the yield, for $\delta=0.01$.}
\label{fig:14}
\end{figure}

\begin{figure}[!t]
\centering
\centerline{\includegraphics[width=4in]{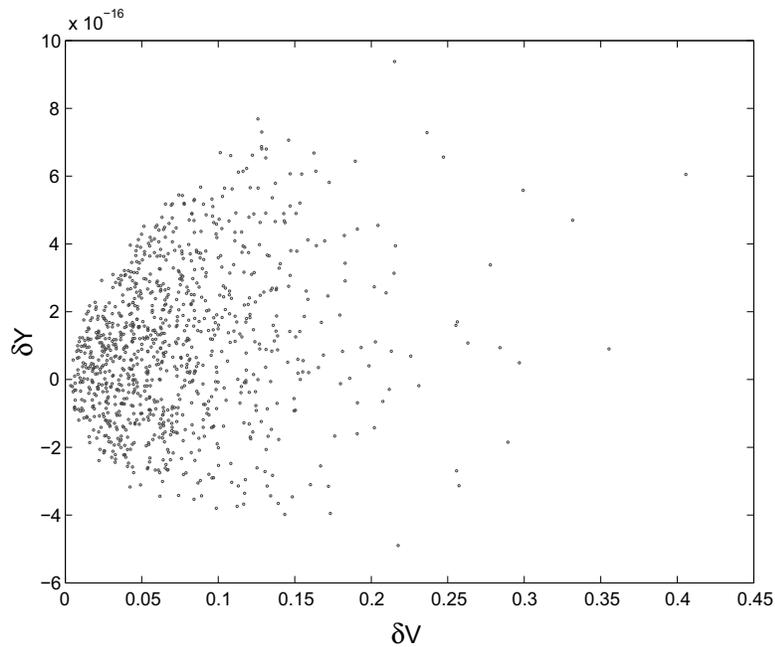}}
\caption{The scatter-plot of $\left(\delta V,\delta Y\right)$ for $\delta=0.01$, $N = 12$ and $M = 8$.}
\label{fig:15}
\end{figure}

Since small changes in $\delta$ result in large changes in the average of the deviation of the yield from its optimum (or equivalently to large deviations in the yield sensitivity) we may expect also the distribution of the yield to be very wide. The scale in the graphs describing those quantities suggest that if we would like to obtain a distribution, it would be reasonable to obtain the distribution of the logarithm of the yield rather than the distribution of the yield itself. Furthermore, after obtaining the distribution of $\delta V$ and the distribution of the logarithm of the yield, it would be interesting to ask if somehow the deviation from the constraints is related to the deviation in the yield. It is natural to assume that the interpolation constraint reduces the yield considerably. Therefore we may expect that the larger $\delta V$ is the larger is the yield. To check our first hypothesis we present first the distribution of the logarithm of the yield (Fig.~\ref{fig:14}). To check our second hypothesis, we present in Fig.~\ref{fig:15} the scatter-plot of $\delta V$ and $\delta Y$ for a fixed $\delta$. Namely, we chose a thousand sets $\{ \delta_m \}$ from the Gaussian distribution with standard deviation  $\delta$. For each set we obtain $\delta V$ and $\delta Y$ and present the pair as a point in the $\delta V - \delta Y$ plane. Both graphs (Figs. \ref{fig:14} and \ref{fig:15}) are presented for $\delta = 0.01$, $N = 12$ and $M = 8$.
We see first that indeed, the distribution of the logarithm of the yield is narrow and resembles a Gaussian. We also see that increasing $\delta V$ tends to increase $\delta Y$, with obvious tendency to have more positive $\delta Y$'s than negative ones. Therefore, the initial insight that relaxing the constraints will result in higher yields seems to be correct. This is actually the reason why the average $\left\langle {\delta Y} \right\rangle $ is always positive.

\subsection{Sensitivity of the number of superoscillations}
\begin{figure}[!t]
\centering
\centerline{\includegraphics[width=4in]{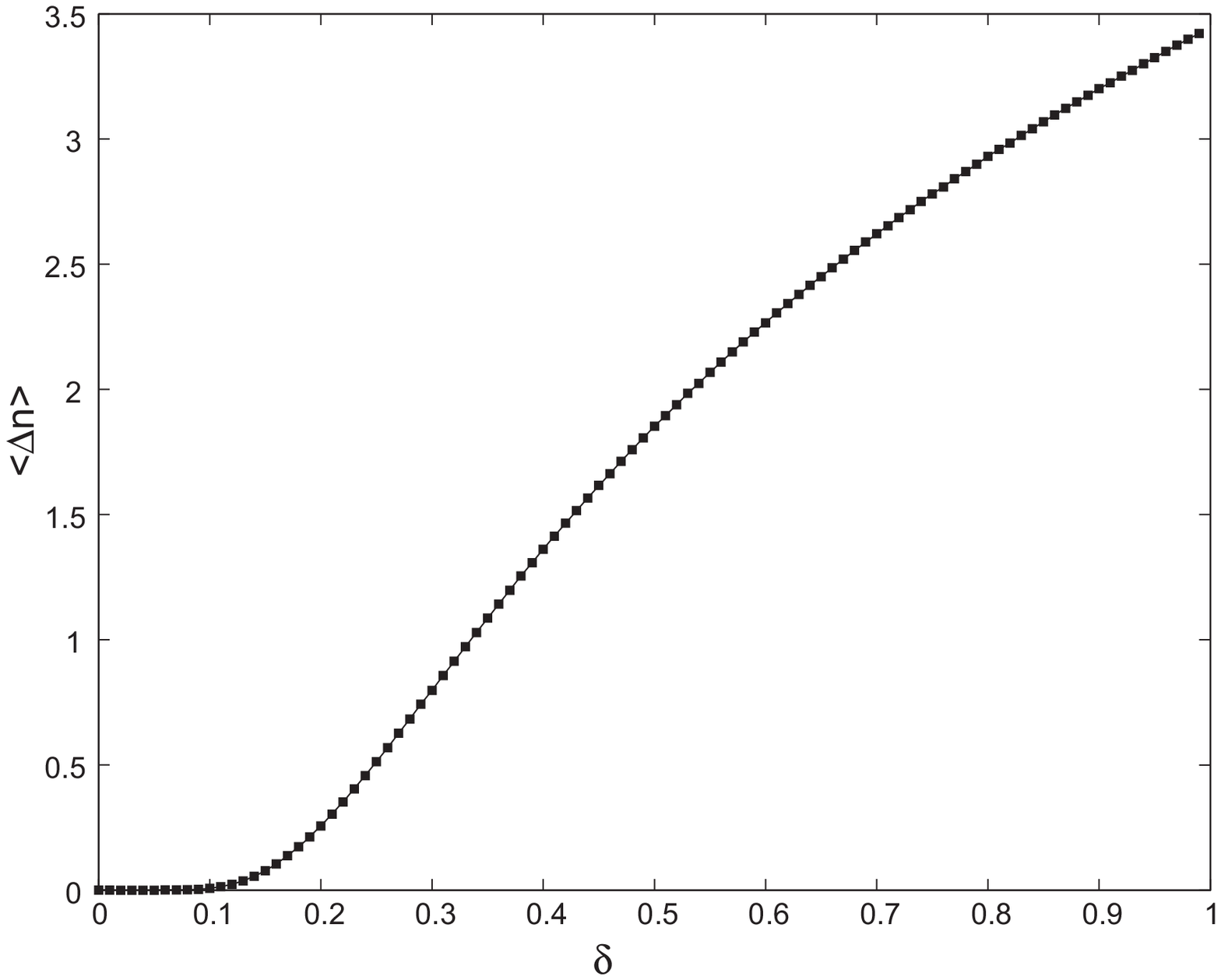}}
\caption{The average number of missing oscillations $\left\langle {\Delta n} \right\rangle$ as a function of of the error level $\delta $ for $N = 15, M = 10$.}
\label{fig:16}
\end{figure}

So far, we have characterized the departure from strict interpolation by the single parameter $\delta V$. It can occur, however, that while the departure from strict interpolation, as expressed by $\delta V$, is not that small, the signal still superoscillates with the prescribed number of oscillation in the appropriate interval. The opposite is also possible. Namely, although $\delta V$ is relatively small, some oscillations may still be lost in the appropriate interval. With that in mind, we count the number of oscillations in the interval $(-a,a)$, and denote the number of missing oscillations in this superoscillation interval by $\Delta n$ (in all the examples we used $a = 1$). In Fig.~\ref{fig:16} we present the average of $\Delta n$ as a function of $\delta$ for the case $N = 15,M = 10$. As expected, the average depart from zero slowly, and stays fairly small even for $\delta$'s that are not extremely small.
The standard deviation of $\Delta n$ as a function of $\delta$ is also a useful tool for assessing the loss of oscillations as a function of the error level $\delta$, and we therefore present it below in Fig.~\ref{fig:17}.
\begin{figure}[!t]
\centering
\centerline{\includegraphics[width=4in]{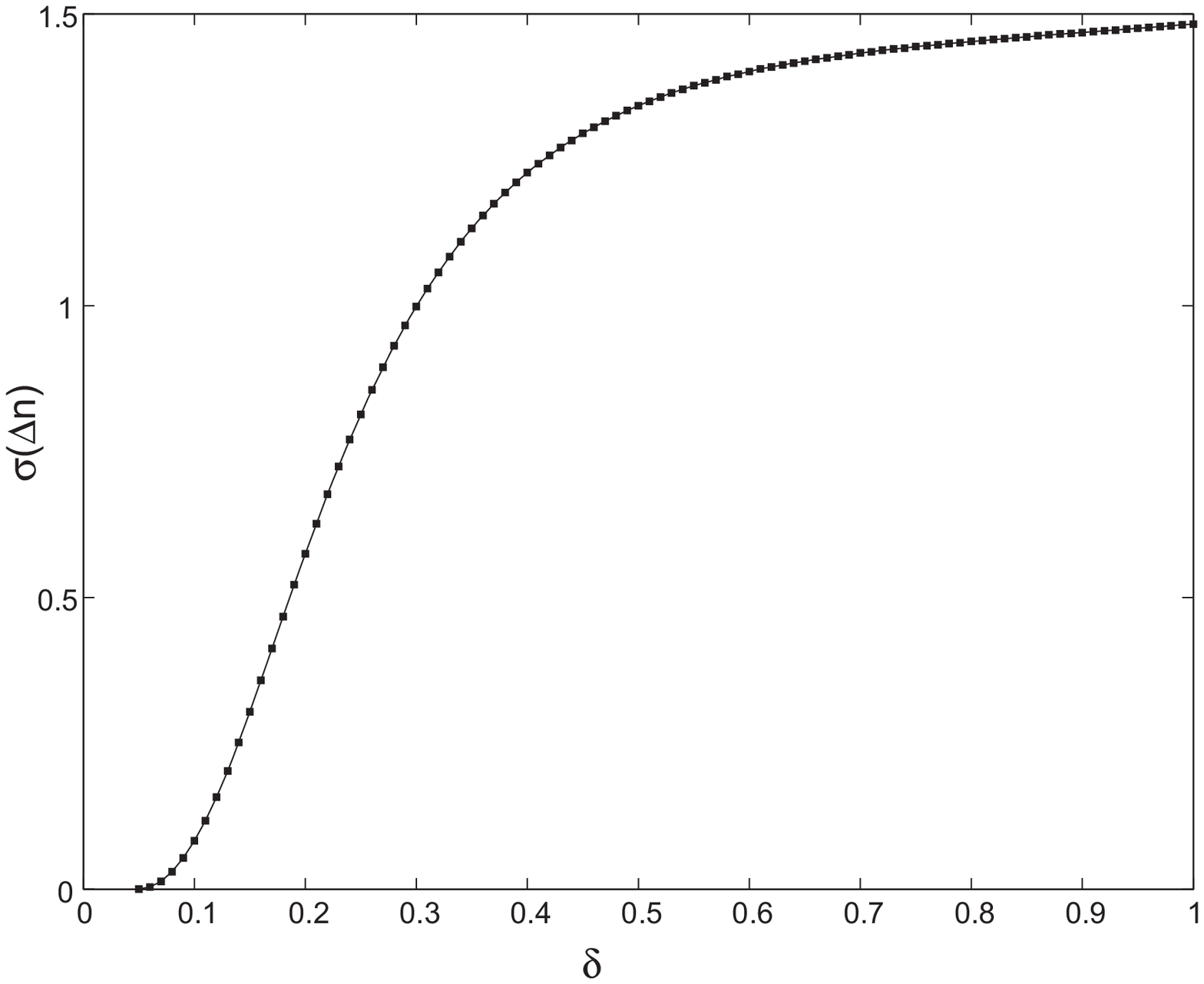}}
\caption{The standard deviation of the number of missing oscillations, as a function of $\delta$ for $N = 15,M = 10$.}
\label{fig:17}
\end{figure}

\section{Summary and Conclusions}
To conclude, we have considered the effect of errors in the Fourier coefficients defining the signal on the superoscillating yield and signal quality of interpolated superoscillations. Such errors are abundant in any physical system that may use superoscillations and therefore analyzing it is of high importance both theoretically and from the point of view of applications.

We considered in detail two scenarios. In the first scenario, we analyzed deviations within the subspace of exactly interpolating signals from that signal that minimizes the signal energy per period. We have also shown that such a signal is very close to the signal of optimal yield within this subspace. The second scenario is where errors in the Fourier coefficients of the low frequency components can violate the interpolation condition and thus can reduce the signal quality in the superoscillation portion. Such errors affect also the yield. To quantify these effects we define the Interpolation-sensitivity $\delta V$ and Yield-sensitivity $\delta Y$. Estimates on both sensitivities as a function of the number of the Fourier components and the standard deviation describing the uncorrelated errors in each of the Fourier coefficients have been obtained. We have also counted the number of missing oscillations $\Delta n$ due to those errors. It is found that an error level $\delta$ in the Fourier components $\left\{A_m\right\}$ which is not too small still ensures superoscillation and can result in a yield which is larger than the optimal one under the interpolation constraint. Namely, a slight relaxation of the strict constraint may still keep the required number of superoscillations yet result in an increased yield which can be a real advantage in application.

An important bottom line is that although generating optimized superoscillations may be a challenging task, as it typically requires very high machine precision, once attained they are no longer very sensitive to error or noise, and can thus be realized in physical systems where noise is abundant. Another important conclusion is that compromising the shape of the function in the superoscillatory region (such as the regularity of the oscillations and the location of their maxima/minima) can result in an improved superoscillatory yield, which could be advantageous.


\end{document}